# ARTICLE

# Diffusion of individual nanoparticles in cylindrical diatom frustule

Naoki Tomioka,*[a] Yusaku Abe [a] and Yu Matsuda *[a]

Diatoms are characterised by silica cell walls (frustules), which have highly ordered micro-/nano-structures. As the synthesis of such structures remains challenging, diatom frustules offer a promising alternative to conventional porous particles in micro-/nano-engineering. In particular, for applications in drag delivery systems, biosensors, and filters, an understanding of particle motion inside diatoms is of great impotance. In this study, we investigated nanoparticle (NP) motions inside diatom frustules using the single particle tracking (SPT) method. For these measruements, the diameter of the NP was about one-tenth smaller than that of the frustule. Inside the frustule, the diffusion motions of the NPs were suppressed, but this suppression was weakened near the exit of the frustule. Moreover, diffusion anisotropy between the axial and radial directions of the frustule was observed. This anisotropy is difficult to detect with ensemble methods; thus, the SPT method is a powerful approach for investigating NP motions in frustules.

## Introduction

Diatoms, single-celled algae found in the oceans, lakes, and waterways, [1, 2] have received considerable attention in various fields. For example, investigations of diatom biology, ecology, systematics, and phylogeny are continuing concern because they play important roles in ocean food webs and biogeochemical cycles, [3] and they contribute more than 20% of annual global carbon fixation on Earth. [3, 4] Since diatoms are used as indicators of water quality, [2, 5, 6] the identification methods of diatoms are also actively studied.

From an engineering point of view, diatoms have a unique porous silica cell wall with highly ordered micro-/nano-structures, called frustules, and high lipid content, which is of interest for micro–/nano– and metabolic engineering applications. [3, 7-15] Here, we focus in particular on porous silica cell walls with ordered structures that are difficult to fabricate with high production rates using current engineering technologies. Thus, diatoms are a promising alternative to mesoporous silica particles as drug delivery systems, [10, 16-20] biosensors, [8, 21-25] and filters. [26-28] These applications depend on structural properties such as large surface areas, large pore volumes, fixed pore sizes, and high porosities. The morphological properties of frustules, such as shape, pore size, and porosity, were investigated using a scanning electron microscopy (SEM), transmission electron microscopy (TEM), and atomic force microscopy. [2, 11, 26] Chemical analyses of frustules have been conducted using X-ray fluorescence and X-ray diffraction. [28] Fluorescence spectroscopy has been used to investigate the photo-chemical and physical properties of frustules. [22-25] The mass transport properties in frustules, such as diffusion coefficient and adsorption rate, have been investigated using high performance liquid chromatography [18] and fluorescence correlation spectroscopy (FCS). [29] These measurements focus on the ensemble−averaged behaviour of particles/molecules in frustules. The behaviour of individual particles/molecules is also informative for further improvement of the application of diatoms in various fields. From this perspective, the images of the adsorption of nanoparticles (NPs) on the surfaces of frustules have been obtained by using SEM [30-32] and TEM. [33] The binding property of Immunoglobulin G to frustule has been visualised using fluorescent microscopy using fluorescent dye. [17] These studies reveal the static properties of individual particles/molecules, but the studies on dynamic properties is required to further promote the application studies of diatoms.

Here, we focus on single−particle/molecule tracking (SPT/SMT) method [34-37] to investigate the dynamic properties of particles/molecules. SPT/SMT is a powerful microscopy technique that measures the motion of individual particles/molecules using a fluorescence microscope. This method has been applied to investigate particle/molecule motions in porous materials, [38, 39] such as metal–organic framework crystals, [40] silica particles, [41] fluid catalytic cracking particles, [42] and monolithic silica columns. [43] In these studies, confined diffusion motions of particles/molecules and deviations from the Gaussian distribution of the displacement probability distribution have been observed. Moreover, the motions of individual particles/molecules, including adsorption and desorption processes, have been successfully detected.

In this study, we applied SPT to investigation of the motions of NPs in the centric diatom *Aulacoseira granulata* (Ehrenberg) Simonsen (syn. *Melosira granulata*), [44] which has a cylindrical

[a.] *Department of Modern Mechanical Engineering, Waseda University, 3-4-1 Ookubo, Shinjuku-ku, Tokyo, 169-8555, Japan*
*Corresponding author: N.T.: a9yiwsm3@toki.waseda.jp and Y.M.: y.matsuda@waseda.jp

shape. The diffusion coefficients of the NPs inside the frustule were investigated. The moment scaling spectra (MSS) analysis [45, 46] was conducted to characterise the NP motions. The distribution of the relative angle between the axial direction of the cylindrical frustule and the direction of particle motion was also analysed.

## Materials and methods

### Sample preparation

As a sample diatom, we adopted *Aulacoseira granulata* (Ehrenberg) Simonsen (NIES-3951), which was provided by the National Institute for Environmental Studies (NIES Collection, Tsukuba, Japan) through the National BioResource Project (NBRP) of the Ministry of Education, Culture, Sports, Science and Technology (MEXT), Japan. The sample diatoms were cleaned to remove organic matters using the following procedure. [47] First, a bleach solution (Pipe-Unish, Johnson Company, Ltd., Japan) containing hypochlorite, sodium hydroxide, and surface–active agent was added to the diatom suspension at a 1: 1 volume ratio. The resulting suspension was stirred gently and allowed to stand for about 20 min. Second, pure water was added to the suspension and centrifuged using a centrifuge (CF18RS, Eppendorf Himac Technologies Co., Ltd., Japan) at 2500 rpm for 10 min. The frustules were obtained as sediments. The obtained frustules were resuspended in pure water. This process was repeated three times. Finally, the frustules were dried and cleaned again using an oxygen plasma cleaner (PC-400T, STREX, Inc., Japan). The SEM image of the frustule coated with Pt is shown in Fig. 1. As shown in the figure, the frustule was cylindrical in shape with a diameter of about 4 μm. Furthermore, highly ordered pores with diameters of approximately 0.5 μm were observed on the surface.

### SPT measurements

We measured particle motions using the SPT method. As probe particles, florescent polystyrene NPs (Fluoro-Max Red R25,

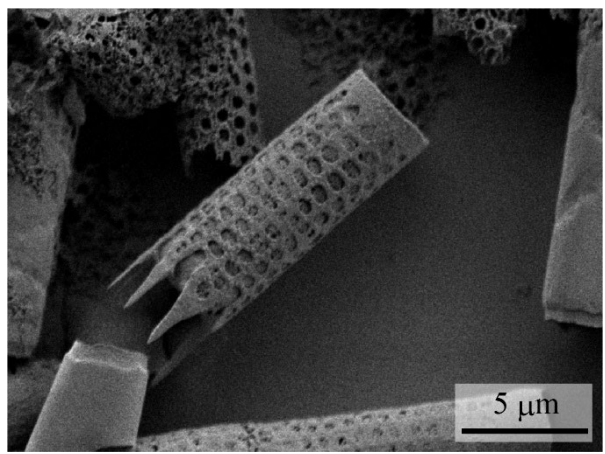

Fig. 1  SEM image of *Aulacoseira granulata* (Ehrenberg) Simonsen (NIES-3951)

Thermo Fisher Scientific Inc., USA) were used. The manufacturer reported that the density of the NP was 1.05 g/cm$^3$. The characteristics of the NP were measured using a nanoparticle analyser (SZ-100, Horiba, Ltd., Japan). In this analyser, the particle diameter was measured using dynamic light scattering (DLS) and the zeta potential of particle was measured using laser doppler electrophoresis. The arithmetic mean and cumulant diameters were measured as 24.1 nm and 21.2 nm, respectively. Thus, the diameter of the NP was about one-tenth that of the frustule. The diffusion coefficient in bulk pure water was $2.26 \times 10^{-11}$ m$^2$/s and the zeta potential was 0 mV. The NP suspension, which was provided as 1 % solids, was diluted 100 times with pure water. Subsequently, the frustules were immersed in the aqueous NP suspension. The resulting suspension was poured into a custom-made silicone-rubber cell placed on a cover glass plate (Thickness No.1, Matsunami Glass Co. Ltd., Japan). The diameter and depth of the silicone-rubber cell were 6 mm and 5 mm, respectively. The motions of the individual NPs were measured using an inverted fluorescence microscope (IX-73, Olympus Co. Ltd., Japan) in the same manner as previously described. [43, 48] We used a confocal scanner unit (CSU-X1, Yokogawa Electric Co. Ltd., Japan) with a zoom lens of 2.0× to obtain confocal images and an oil-immersion objective lens of 100×, NA = 1.45, and WD = 0.13 mm (UPLXAPO100XO, Olympus Co. Ltd., Japan). The position of the objective lens was controlled using a piezo actuator (P-725K, Physik Instrumente GmbH & Co. KG, Germany). The focal plane was set 5 μm above the surface of the cover glass to avoid the hindering of the diffusion due to the surface. [49, 50] As an illumination light source, a solid-state laser with an emission wavelength of 488 nm (OBIS488LS, Coherent CO. Ltd., USA) and output power of 120 mW was used. Fluorescence emitted from the NPs was filtered by a bandpass filter (FF01-565/133-25, Semrock, USA) to eliminate the illumination light. This fluorescence was recorded using an electron-multiplying charge-coupled device (EMCCD) camera (C9100-23B, ImagEM X2, Hamamatsu Photonics Co. Ltd., Japan) at 20 or 33 frames per second (fps) (exposure time: 50 ms or 30 ms ). In the obtained images, the one pixel corresponded to an actual length of 0.08 μm. All SPT measurements were conducted at the temperature of 25 °C in this study.

### Data analysis of trajectories

The obtained SPT data were analysed as described in previous studies. [43, 48] Here, we briefly introduce the analysis method; detailed explanations can be found in the literature. [46] The particle positions and trajectories were extracted using ParticleTracker, [46, 51] where the parameters "radius", "cutoff", and "Per/Abs" were respectively set to 5 px, 0, and 1. The position vector of a nanoparticle at the $n$-th frame is represented by $\mathbf{r}(n)$, where $n = 0, 1, 2, \cdots, N$ and $N$ is the total number of frames tracking the particle. The diffusion coefficient $D$ was calculated based on the mean squared displacement (MSD). The moment of displacement with order $\nu$ is written as

$$\delta^\nu(\Delta n) = \frac{1}{N - \Delta n} \sum_{n=0}^{N-\Delta n-1} \|\mathbf{r}(n + \Delta n) - \mathbf{r}(n)\|^\nu, \quad (1)$$

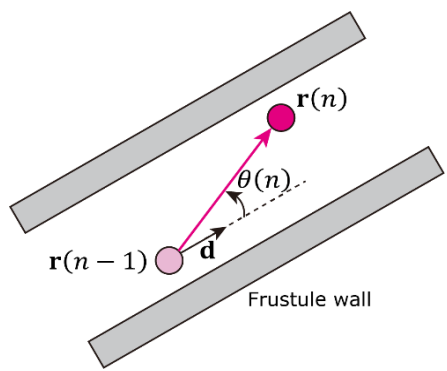

Fig. 2 Schematic of angle between central axis of frustule and direction of particle motion.

where $\Delta n$ is the frame shift and $\|\blacksquare\|$ denotes the Euclidean norm. When $\nu = 2$, Eq. 1 is known as the MSD. The relation between the diffusion coefficient $D$ and MSD with the shift time $\tau = \Delta n \Delta t$ is represented as

$$\delta^2(\Delta n) = 4D\tau, \quad (2)$$

where $\Delta t$ is the time lag, which corresponds to the frame interval. The moment of $\delta^\nu(\Delta n)$ also depends on $\tau$ as $\delta^\nu(\Delta n) \propto \tau^\gamma$, where $\gamma$ is the scaling coefficient determined by least-squares regression of the $\log \delta^\nu(\Delta n)$ versus $\log \tau$ plot. Then, the obtained $\gamma$ and $\nu$ plot is called the MSS. [45] It is known that the slope of MSS $S_{MSS}$ indicates the type of particle motion: $0 < S_{MSS} < 0.5$ for supressed/confined diffusion, $S_{MSS} \sim 0.5$ for free diffusion, and $0.5 < S_{MSS} < 1$ for super/drifted diffusion including the Lévy flight.

The cylindrical shape of the frustule is expected to affect particle motion. Therefore, we investigated the distribution of the angle $\theta$ between the axial direction of the cylindrical frustule and the direction of particle motion, as shown in Fig. 2. The angle $\theta$ is defined as

$$\theta(n) = \cos^{-1}\left[\frac{\Delta \mathbf{r}(n) \cdot \mathbf{d}}{\|\Delta \mathbf{r}(n)\|}\right], \quad (3)$$

where $\mathbf{d}$ is a direction unit vector along the axial direction of the cylindrical frustule, $\Delta \mathbf{r}(n) = \mathbf{r}(n) - \mathbf{r}(n-1)$, $n > 2$, and the operator "$\cdot$" indicates the dot product. The particle with larger displacement will be affected by the frustule wall. Then, the following $\alpha - \beta$ space scatter diagram, which is used to evaluate the diffusion anisotropy, [52] is considered:

$$\begin{bmatrix} \alpha_n \\ \beta_n \end{bmatrix} = \frac{\|\Delta \mathbf{r}(n)\|}{\Delta t}\begin{bmatrix} \cos \theta(n) \\ \sin \theta(n) \end{bmatrix}. \quad (4)$$

Note that $\|\Delta \mathbf{r}(n)\|/\Delta t$ is not the ordinary speed of the particle. The radius of gyration tensor $\mathbf{R}_g$ is used to characterize the shape of the scatter diagram [53] as

$$\mathbf{R}_g^2 = \begin{bmatrix} \frac{1}{N}\sum_{i=1}^{N}(\alpha_i - \langle\alpha\rangle)^2 & \frac{1}{N}\sum_{i=1}^{N}(\alpha_i - \langle\alpha\rangle)(\beta_i - \langle\beta\rangle) \\ \frac{1}{N}\sum_{i=1}^{N}(\alpha_i - \langle\alpha\rangle)(\beta_i - \langle\beta\rangle) & \frac{1}{N}\sum_{i=1}^{N}(\beta_i - \langle\beta\rangle)^2 \end{bmatrix}. \quad (5)$$

The radius of gyration tensor is symmetric. The eigen vector with a larger eigen value points in the direction in which the span of the scatter diagram is maximised, whereas that with a smaller eigen value points in the direction in which the span is minimised. For normal diffusion, the ratio of the eigen values is unity because there is no special direction in the motion.

## Results and discussion

### Individual particle motions

A typical particle motion is shown in Fig. 3. The trajectory in the figure was captured at 33 fps and is overlaid on a white-black image of the frustule. In this measurement, the particle was initially inside the frustule and then went out of it. We analysed the particle motion according to the location of the particle: inside the frustule, inside the frustule near the exit, or outside the frustule near the exit. The diffusion coefficient and MSS slope, $S_{MSS}$, inside the frustule were calculated as $7.0 \times 10^{-13}$ m$^2$/s and 0.2, respectively. Compared with the diffusion coefficient in bulk solution of $2.26 \times 10^{-11}$ m$^2$/s, the diffusion coefficient inside the frustule was highly suppressed. This effect was also supported by the small $S_{MSS}$ of $0.2$. The diffusion coefficient and MSS slope $S_{MSS}$ inside near the frustule exit were calculated as $1.3 \times 10^{-12}$ m$^2$/s and 0.4, respectively. The diffusion coefficient and $S_{MSS}$ were larger than those inside the frustule. The diffusion coefficient and MSS slope $S_{MSS}$ outside near the exit were calculated as $3.7 \times 10^{-12}$ m$^2$/s and 0.5, respectively. Thus, as the particle approached the exit, the diffusion coefficient increased and the diffusion mode changed from confined motion with $S_{MSS} = 0.2$ to Brownian motion with $S_{MSS} = 0.5$. In our measurements, no events were observed in which the NPs entered and exited the frustule through the pores on the surface.

### Statistical analysis of particle motions

In this study, we analysed a total of 101 trajectories inside the frustule, inside near the frustule exit, and outside near the frustule exit, where the trajectories with more than six frames were considered for reliable analysis. Fig. 4 (a) and (b) show the histograms of the diffusion coefficient and MSS slope, respectively, for these trajectories. The number of bins in the histograms was determined by the Sturges's rule. As shown in Fig. 4 (a), the diffusion coefficients of most NPs were about one-tenth that in the bulk solution. The histogram of the MSS slope (Fig. 4(b)) indicates that about half of the NPs moved with almost free diffusion motion. This behaviour differs from the SPT results for NPs in a monolithic silica column in our previous study,[43] where many adsorbed NPs were observed owing to the presence of small pores (bimodal pore size distribution with pores of 52 nm and 2 μm).

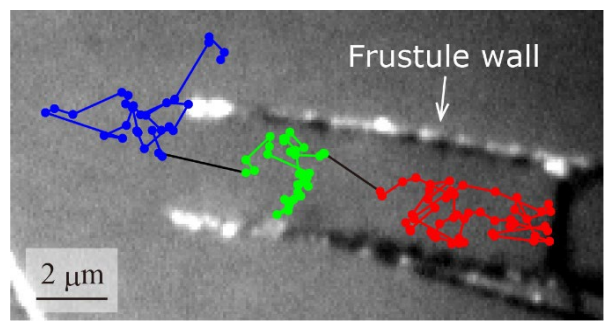

Fig. 3 Typical observed particle motion. The trajectory is coloured by the location of the particle: inside the frustule, inside near the exit, and outside near the exit.

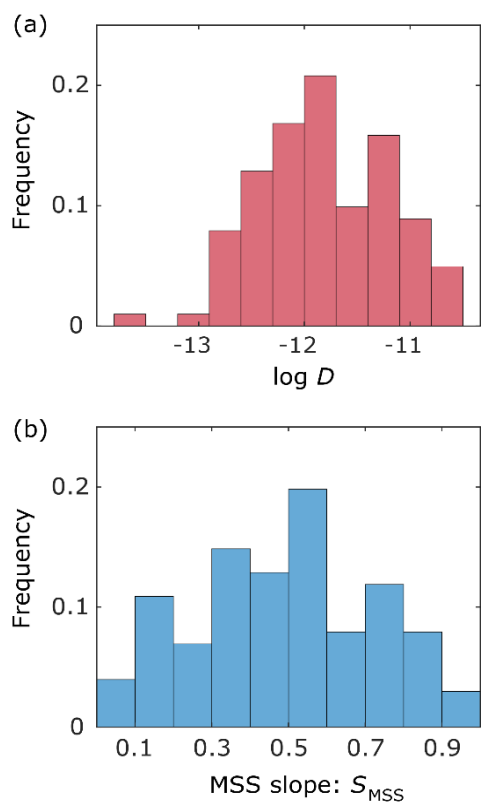

Fig. 4 Statistical properties of measured particle trajectories. (a) Histogram of the diffusion coefficient. The horizontal axis is logarithm of the diffusion coefficient. (b) Histogram of the MSS slope.

**Angle of particle motions**

As a further analysis, we considered the distribution of the angle $\theta$ between the axial direction of the cylindrical frustule and the direction of particle motion following Eqs. (4) and (5). Fig. 5 shows the obtained scatter diagram, where the axis units are μm/s. It is noted again that these are not the ordinary speeds of the particles. The diagonal elements of the radius of gyration tensor $\mathbf{R}_g$ were 73.9 and 58.4, respectively. The off-diagonal element was about one-hundredth of the diagonal element; thus, $\mathbf{R}_g$ was almost a diagonal tensor. The ratio of the eigen values was 1.3, indicating an anisotropy diffusion. This result indicates that an NP was likely to move a longer distance in the axial direction of the cylindrical frustule and a shorter distance in the radial directions during the time lag $\Delta t$. This anisotropy is difficult to detect with ensemble methods such as FCS and DLS, thereby demonstrating the suitability of SPT for such analyses.

Next, the displacement probability distributions (DPDs) of the NPs were considered. For Brownian motion, the DPD is expressed as a Gaussian distribution. The DPDs along the axial and radial directions are shown in Fig. 6. As shown in the figure, the DPD in the axial direction of the frustule deviated from that in the radial direction and from the Gaussian distribution.

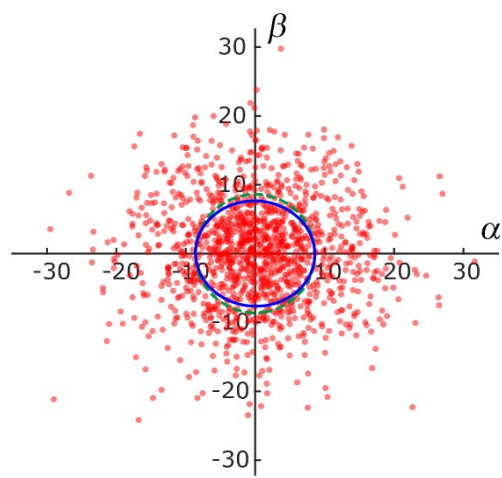

Fig. 5 Scatter diagram for angle of particle motion relative to axial direction of cylindrical frustule. The unit of the axes is μm/s. The blue solid line is the ellipse obtained from the radius of gyration tensor. The green dashed line is a circle for comparison.

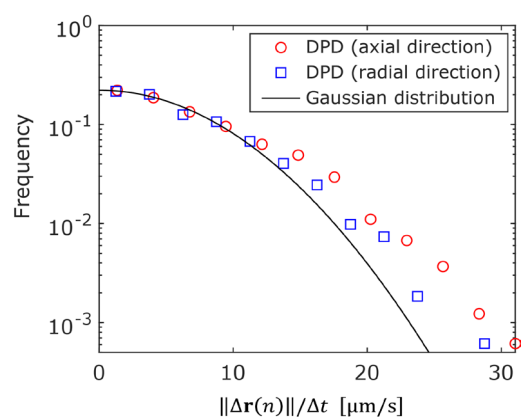

Fig. 6 Displacement probability distributions in axial and radial directions of frustule. The solid line is a Gaussian distribution.

## Conclusions

In this study, we investigated the motions of nano-particles (NPs) in a diatom frustule with a cylindrical shape using the single particle tracking (SPT) technique, which enabled the investigation of individual NP motions inside the frustule. It is found that the diffusion motions of the NPs inside the frustule were suppressed; however, this suppression became weaker near the exit of the frustule. In our measurement, no events corresponding to NPs entering and exiting the frustule through the pores on the surface were observed. Therefore, further investigation is required to detect and analyse such rare events. The analysis based on the moment scaling spectra (MSS) indicated that fewer NPs were adsorbed in the frustule than in a monolithic silica column, likely because the free volume of the frustule was larger than that of the silica column. Moreover, evaluation of the distribution of the angle between the axial direction of the cylindrical frustule and the direction of particle motion revealed that a NP was likely to move a longer distance in the axial direction of the cylindrical frustule than in the radial direction during the time scale of the SPT measurement. That is, the anisotropy of the diffusion motion was observed by using the SPT method. It is also found that the displacement probability distribution (DPD) in the axial direction deviated from a Gaussian distribution. The SPT method is a powerful tool for extracting the anisotropy of diffusion motion, which is difficult to detect using ensemble averaged methods such as fluorescence correlation spectroscopy (FCS) and dynamic light scattering (DLS).

## Author contributions

Naoki Tomioka: Data curation, Formal analysis, Investigation, Methodology, Software, Validation, Visualization, Writing – original draft, Writing – review & editing.
Yusaku Abe: Investigation, Methodology, Software, Validation
Yu Matsuda: Conceptualization, Data curation, Funding acquisition, Investigation, Methodology, Project administration, Supervision, Writing – original draft, Writing – review & editing

## Conflicts of interest

There are no conflicts to declare.

## Acknowledgements

The authors thank H. Imai and R. Numajiri for assistance with SPT measurements. This work was partially supported by the JSPS, Japan Grant-in-Aid for Scientific Research (B), No. 22H01421.